\documentclass[12pt,a4paper,%
 aip,
 jmp,%
 amsmath,amssymb,
preprint,%
]{revtex4-1}

\draft 

\pdfoutput=1
\usepackage{bbold}
\usepackage{graphicx}
\usepackage{bm,amsmath,amssymb}
\usepackage{pdfsync}
\usepackage{enumerate}
\usepackage{hyperref}

\newcommand\eq[1]{\begin{eqnarray}#1\end{eqnarray}}

\newcommand{\diff}{\text{d}}
\newcommand{\ad}{\text{ad}}

\newcommand{\EXP}{\text{EXP}}

\newcommand{\dexpR}[1]{\left( \frac{e^{\ad #1}-1}{\ad #1} \right)}
\DeclareMathOperator{\tr}{Tr}
\begin{document}
\title{Infinite symmetry on the boundary of $SL(3)/SO(3)$}
\author{Heikki Arponen}
\noaffiliation
\affiliation{Currently unaffiliated}
\email{heikki.a.arponen@gmail.com}
\homepage{https://sites.google.com/site/heikkialeksiarponen/}
\date{\today}
%

%
\begin{abstract}
Asymptotic symmetries of the five dimensional noncompact symmetric space $SL(3,\mathbb{R})/SO(3,\mathbb{R})$ are found to form an infinite dimensional Lie algebra, analogously to the asymptotic symmetries of anti-de Sitter/hyperbolic spaces in two and three dimensions. Possible generalizations of the AdS/CFT correspondence and gauge/gravity dualities to such a space is discussed.
\end{abstract}
%
\maketitle
\section{Introduction}
%
%

Asymptotic symmetries in gauge theories are defined as transformations that leave the field configurations asymptotically invariant. In gravity, a standard example of asymptotic symmetries is the three dimensional (Euclidean) Anti-de Sitter space, which employs the idea of conformal compactification due to Penrose\cite{penrose}. The $AdS_3$ space can be defined as a space of constant negative curvature with a metric
\eq{ds^2 = dr^2+ \sinh^2 (r)d\Omega^2, \label{adsmetric}}
where $d\Omega^2$ is the metric on the unit sphere.\footnote{The Euclidean AdS spaces are in fact just hyperbolic spaces, but we choose to use the $AdS$ terminology due to it's relation to field theory.} Evidently this metric is singular in the limit $r\to \infty$ and therefore not well defined on the boundary. One can extend the metric to include the boundary by redefining
\eq{d  \tilde{s}^2 = f^2 ds^2,}
where the function $f$ is defined up to an asymptotic equivalence class $f\sim e^{-r}$. Since $f$ is otherwise arbitrary, a transformation which maps $d\Omega^2 \to \Lambda d\Omega^2$, where $\Lambda$ is an arbitrary positive function of $\theta,\phi$, is an \emph{asymptotic symmetry}\footnote{Note that sometimes the notion of an asymptotic symmetry is used to refer to asymptotic symmetries of gravity theories that are only asymptotically $AdS$ but otherwise free. Here we make no such distinction} of the above metric. A more rigorous definition can be found in the references \cite{barnich,henneaux,witten}. The asymptotic symmetries of the $AdS_3$ space then form the infinite dimensional Lie algebra of conformal transformations \cite{henneaux,witten}.\\

The idea behind asymptotic symmetries can perhaps be better appreciated by the following example. Consider the $AdS_2$ space (a.k.a. hyperbolic space $\mathbb H^2$) in the Poincar\'{e} upper half plane $\left\{(x,y)|y>0\right\}$ with the metric
\eq{ds^2 =\frac{1}{y^2} \left(dx^2 + dy^2 \right).}
The boundary is now at $y \to 0$. The geodesics are semi-circles with origins on the $x$-axis, i.e. determined by equations of the form $(x-x_0)^2+y^2 = \text{const}$. We can think of the geodesics as trajectories of a system of particles satisfying an equation of motion, which is the geodetic equation. The Lie algebra of isometries is $\mathfrak{sl}(2,\mathbb R)$, generated by the Killing vector fields
\eq{\left\{ \begin{aligned}
l_{-} &= \partial_x\\
h &=x \partial_x +y \partial_y \\
l_{+} &= (x^2-y^2)\partial_x +2 x y \partial_y.
\end{aligned}\label{sl2.generators}
\right. }
These are also symmetries of the "particle system", since they map geodesics to other geodesics. The asymptotic metric can then be obtained e.g. by rescaling $y \to \lambda y$ and taking the limit $\lambda \to 0$, which corresponds to $ds^2 \sim dx^2 /(\lambda y)^2$.
\begin{widetext}
\begin{center}
\begin{figure}[hb]
\includegraphics[scale=.4]{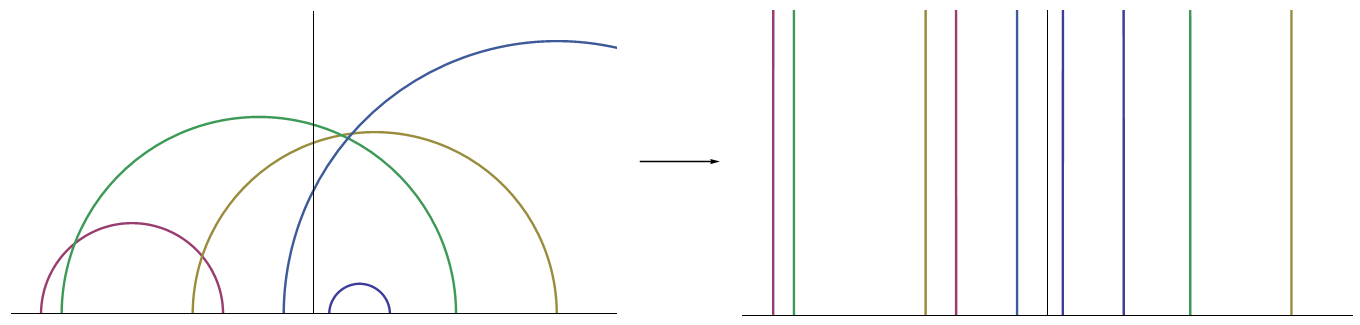}
\caption{A random assortment of geodesics as seen near the boundary in the limit $\lambda \to 0$.}
\label{fig:example}
\end{figure}
\end{center}
\end{widetext}
What happens to the geodesics can be seen in Fig. (\ref{fig:example}), i.e. the geodesics are mapped to vertical lines. Clearly this asymptotic "particle system" is invariant under the asymptotic symmetries $x \to f(x)$. This is of course a drastically simplified example, although the idea remains essentially the same e.g. in the work of Coussaert, Henneaux and van Driel\cite{henneaux.coussaert}, where the authors deduced that the asymptotic dynamics of gravity in three dimensions with AdS boundary conditions is described by the two dimensional Liouville conformal field theory.\\

In the seminal paper \cite{henneaux} Brown and Henneaux (see also \cite{henneaux.coussaert,henneaux.banados}) did not only consider the asymptotic symmetries of the $AdS_3$ metric, but instead a full fledged Hamiltonian formulation of gravity with asymptotically AdS boundary conditions, i.e. that in the limit $r \to \infty$, the metric must approach the solution in eq. (\ref{adsmetric}). They showed, quite surprisingly, that the infinite dimensional Poisson bracket algebra corresponding to the asymptotic symmetries possesses a nontrivial central extension. A little later Ach\'{u}carro and Townsend showed that three dimensional AdS gravity can be expressed as a Chern-Simons theory\cite{achucarro:1987}, and Witten showed that the three dimensional Chern-Simons theory is exactly solvable due to its equivalence to a two dimensional boundary conformal field theory\cite{Witten:1988}. Later it was also shown that the central extension (or rather the central charge) yields the Bekenstein-Hawking entropy of some black holes via Cardy's formula\cite{strominger}. The discovery has lead several authors to conjecture even more explicit dualities between three dimensional gravity theories and boundary conformal field theories, such as in the example mentioned above, and e.g. in higher spin Chern-Simons gravity \cite{gaberdiel,henneaux2}.\\

Another important manifestation of the asymptotic symmetry concept is the AdS/CFT correspondence\cite{witten}, which provides an explicit (although perhaps not unique) relationship between the bulk (AdS) and boundary (CFT) theories. At its simplest form, the bulk theory can be approximated as living on a fixed AdS background. Although the origins of the AdS/CFT correspondence lie in string theory, it should be noted that the asymptotic symmetry concept is quite independent of it. The different approaches may all be seen as realizations of the holographic principle\cite{thooft,susskind}, which conjectures that the degrees of freedom inside a volume are somehow encoded on the boundary of the volume. The principle can be seen as an attempt to resolve the black hole information paradox\cite{susskind}.\\

Yet another side of holography, gauge/gravity duality and asymptotic symmetries may perhaps be seen in the renormalization group. It has been conjectured that the "extra dimension" in the AdS space corresponds to an energy scale of the theory being renormalized. Attempts have been made toward interpreting the (Wilsonian) renormalization group in such a holographic manner, thus combining it with the concept of the holographic renormalization group (see e.g. Heemskerk and Polchinski\cite{heemskerk-polchinski}).\\

The unifying theme throughout the geometric asymptotic symmetry ideas has been the Anti-de Sitter space. The extension to $d>3$ AdS spaces can be carried out by simply replacing the metric on the unit sphere in eq. (\ref{adsmetric}) by a corresponding metric on a $d-1$ dimensional unit sphere. However, the conformal group is finite dimensional in higher dimensions, and therefore hopes of integrability (in the 2D CFT sense) of the boundary (and perhaps the bulk) theory is lost.\\

It is possible to extend the study of asymptotic symmetries to noncompact symmetric spaces. A symmetric space is defined as a homogeneous space $G/K$, where $G$ is a Lie group and $K$ is a maximal compact subgroup invariant under the Cartan involution (see e.g. Helgason\cite{helgason} for the symmetric space construction). The example that will be considered in this work is the space $P(3,\mathbb{R})\doteq SL(3|\mathbb{R})/SO(3|\mathbb{R})$, which is a five dimensional symmetric space of constant negative Ricci scalar curvature\footnote{The notation naturally extends to $P(N,\mathbb{R})\doteq SL(N|\mathbb{R})/SO(N|\mathbb{R})$}. A notable difference to a five dimensional AdS space is that its rank, and therefore the number of "radial" coordinates, is two, whereas the rank of AdS space is one in all dimensions. The boundary of the space is thus divided into different regimes according to which of the radial variables is taken to infinity. It will be shown that at each limit to the boundary the asymptotic symmetries form an infinite dimensional Lie algebra. The situation should be compared to the $AdS$ spaces, where such an infinite symmetry arises only in dimensions two and three. It should be stressed that we consider here only the case of a fixed "background metric" or gauge field. The cases of gauge/gravity theories with $P(3)$ -like asymptotic boundary conditions and various field theories on a fixed $P(3)$ background will be left for future works.

%
%
%
\section{Noncompact symmetric spaces}
Some properties of noncompact symmetric spaces will be discussed in this section. We will concentrate mostly on $P(3,\mathbb{R})=SL(3|\mathbb{R})/SO(3|\mathbb{R})$, although the $P(2,\mathbb R) \simeq AdS_2$ and $AdS_3$ spaces deserve to be mentioned also.\footnote{We stress again that we mean the Euclidean versions of anti-de Sitter spaces.} Specifically, isometries of $AdS_2$ are generated by $\mathfrak{sl}(2|\mathbb{R})$ and isometries of $AdS_3$ by the direct sum $\mathfrak{sl}(2|\mathbb{R}) \oplus \mathfrak{sl}(2|\mathbb{R}) \simeq \mathfrak{so}(1,3|\mathbb{R})$. Much of the formalism below follows the material in the books cited in the references\cite{helgason,gilmore,magnea,fecko}. The book by Gilmore\cite{gilmore} and the review article by Magnea\cite{magnea} should be singled out for their accessibility to physicists. The asymptotic geometry of spaces $P(N,\mathbb R)$ has been studied e.g. by Schr\"{o}der\cite{schroeder}, but the asymptotic symmetries seem to be an original discovery of the present author.\\

\subsection{Metric on a symmetric space}
Let $G$ denote a Lie group with Lie algebra $\mathfrak{g}$. In the following, we consider only $G=SL(N|\mathbb{R})$. To streamline notation, we denote a basis of $\mathfrak{g}$ as $\left\{X_i \right\}$ while simultaneously thinking of the $X_i$ as the matrices of the defining representation, $X_i \in \mathfrak{gl}\! \left( \mathbb{R}^N \right)$.
Another representation frequently needed is the adjoint/regular representation, defined as the map $\ad: \mathfrak{g} \to \mathfrak{gl}(\mathfrak{g})$ of the Lie algebra to $\text{dim}(\mathfrak g)\times \text{dim}(\mathfrak g)$ matrices. Explicitly,
\eq{\ad X (Y) \doteq \left[X ,Y \right].}
We then denote $X(\omega) \doteq \omega^i X_i$ with ${\omega^i} \in \mathbb{R}$. For $\mathfrak{sl}(N|\mathbb{R})$ Lie algebras, the Cartan involution is defined as $\theta(X) \doteq - X^T$. Cartan decomposition of a Lie algebra then consists of writing the algebra in terms of subspaces, $\mathfrak{g} = \mathfrak{p}+\mathfrak{k}$, with $\theta(\mathfrak{k}) = + \mathfrak{k}$ and $\theta(\mathfrak{p}) = - \mathfrak{p}$. The subsets can then be shown to satisfy the relations
\begin{align}
\left[\mathfrak{p},\mathfrak{p} \right]  &\subseteq \mathfrak{k}  &\left( \mathfrak{p}, \mathfrak{p}\right) &>0\\
\left[\mathfrak{p},\mathfrak{k} \right]  &\subseteq \mathfrak{p}, &\left( \mathfrak{p}, \mathfrak{k}\right) &= 0\\
 \left[\mathfrak{k},\mathfrak{k} \right]  &\subseteq \mathfrak{k}  &\left( \mathfrak{k}, \mathfrak{k}\right)&<0,
\end{align}
where the inner product is defined in terms of the Killing form, $\kappa (X,Y) \doteq \langle X,Y\rangle \doteq\tr \left(X Y \right)$. $\mathfrak{k}$ is a maximal compact subalgebra of $\mathfrak g$. A symmetric space can then be defined as
\eq{P \doteq G/K = \EXP(\mathfrak{p}).}
The direct exponentiation of $\mathfrak p$ as above is usually referred to as canonical coordinates of the first kind. Other parametrizations are naturally possible, and in fact we will be resorting to one below. For now however it is useful to consider the canonical coordinates. We define a right invariant Lie algebra valued Maurer-Cartan (MC) form on $P$ as
\eq{\Omega_R (\mathfrak p)  \doteq dp p^{-1} = X_i \mathfrak{R}^i_{\  \mu}\! \left( \mathfrak{p} \right) d\omega^\mu \doteq X_i \otimes \theta^i,\label{MCform}}
where $p \in P$ and we denote the elements/coordinates in $\mathfrak g$ by latin indices $\{i,j, \ldots\}$ and the coordinates on $\mathfrak p$ by greek indices $\{\mu, \nu, \ldots\}$. In the case of the canonical coordinates, the functions $\mathfrak{R}^i_{\  \mu}\! \left( \mathfrak{p} \right)$ can be obtained by the following formula (see e.g. p.120 of Helgason\cite{helgason}):
\eq{\partial_i \exp \left( X(\omega) \right)= X_j \dexpR{X(\omega)}_{\!\!ji}\!\!\!  \exp \left( X(\omega) \right),\label{derivativeexp}}
which results in the expression\footnote{In practice however, it may be more convenient to just exponentiate $\mathfrak p$ in the defining representation, take the derivatives and solve for $\mathfrak{R}$.}
\eq{\mathfrak{R}(X) \doteq \dexpR{X}.}
The metric on $P$ is then defined as
\eq{\diff s^2 \doteq \langle \Omega_R,\Omega_R \rangle = d\omega^\mu \left( \mathfrak{R}^T  \kappa  \mathfrak{R} \right)_{\mu \nu}d\omega^\nu,\label{metric}}
where $\kappa$ is again the Killing form (we will usually denote also by $\kappa$ the matrix with elements $\kappa_{ij} = \kappa\left(X_i,X_j \right)$). Its group of isometries is just the left actions of the Lie group $G$ itself. We are now interested in the asymptotic metric, i.e. the parts that diverge on the boundary, similarly as in the AdS metric of eq. (\ref{adsmetric}).


\subsection{Asymptotic metrics of $P(3,\mathbb{R})$}
We decompose the $\mathfrak{sl}(3,\mathbb R)$ Lie algebra into following subsets:
\eq{\mathfrak{h} &\doteq
    \left(\begin{array}{ccc}
     r^1 & 0 & 0\\
     0 & -r^1\! \!-\! r^2 & 0\\
     0 & 0 & r^2 \\
    \end{array}\right),\\ \mathfrak{k} &\doteq
    \left(\begin{array}{ccc}
     0 & -z & -y\\
     z & 0 & -x\\
     y & x & 0 \\
    \end{array}\right), \\ \mathfrak{p}' &\doteq
    \left(\begin{array}{ccc}
     0 & \omega^6 & \omega^7\\
     \omega^6 & 0 & \omega^8\\
     \omega^7 & \omega^8 & 0 \\
    \end{array}\right).}
Here $\mathfrak{k} = \mathfrak{so} (3, \mathbb R)$ is the maximal compact subalgebra, $\mathfrak h$ is the Cartan subalgebra and $\mathfrak p'$ are the rest of the noncompact generators. Note that the Killing form is not diagonal in this basis, indeed
%
%
\eq{\kappa = \left(\begin{tabular*}{0.43\textwidth}{@{\extracolsep{\fill}}cccccccc}
     2 & 1 &   &   &   &   &   &  \\
     1 & 2 &   &   &   &   &   &  \\
       &   & -2&   &   &   &   &  \\
       &   &   & -2&   &   &   &  \\
       &   &   &   & -2&   &   &  \\
       &   &   &   &   & 2 &   &  \\
       &   &   &   &   &   & 2 &  \\
       &   &   &   &   &   &   & 2\\
\end{tabular*}\right).}
%
%
It turns out to be more useful to use the Cayley parametrization of the elements in the maximal compact subgroup $SO(3,\mathbb R)$ by defining
\begin{widetext}
\eq{k &&\doteq \left(1 - \mathfrak{k} \right) \left(1 + \mathfrak{k} \right)^{-1}\nonumber \\  &&= \frac{1}{1+x^2+y^2+z^2} \left(
\begin{array}{ccc}
 1+x^2-y^2-z^2 & -2 x y+2 z & 2 (y+x z) \\
 -2 (x y+z) & 1-x^2+y^2-z^2 & 2 (x-y z) \\
 -2 y+2 x z & -2 (x+y z) & 1-x^2-y^2+z^2
\end{array}
\right).}
\end{widetext}
An element $p \in P(3,\mathbb R)$ can then be parametrized as\cite{magnea}
\eq{p \doteq k h k^T,}
where
\eq{h \doteq \EXP (\mathfrak h) =
\left(
\begin{array}{ccc}
 e^{r^1} & 0 & 0 \\
 0 & e^{-r^1-r^2} & 0 \\
 0 & 0 & e^{r^2}
\end{array}
\right).}
Such general polar decompositions are always possible
for noncompact symmetric spaces $SL(N)/SO(N)$ (see e.g. \cite{hermann1966}, Ch.6). The coordinates of the symmetric space are simply denoted as $\{\omega^1, \ldots. \omega^5 \} = \{ r^1, r^2, x,y,z \}$. It is now evident that the noncompact coordinates $r^i$ can be thought of as radial coordinates on the symmetric space. The number of radial coordinates on $P(N,\mathbb R)$ is equal to $N=\text{dim}(\mathfrak h)=\text{rank}(\mathfrak g)$. We will consider only the cases $r^1 \to \infty$ and $r^1 ,r^2 \to \infty$ since the other cases are very similar and can be obtained by permutations of $\{x,y,z\}$. The weights of the defining representation are $\{\lambda_i\}= \left\{ r^1, r^2, -r^1-r^2 \right\}$ and the roots (weights of the adjoint representation) are the differences $\left\{ \lambda_i - \lambda_j \right\} = \left\{ r^1-r^2, 2 r^1+r^2, r^1+2 r^2 \right\}$ and their negatives. This observation leads also to a simplification of the asymptotic one forms: from the considerations above, we know that the derivatives of $p$, i.e. the matrix $\mathfrak R$, are represented as functions of the adjoint representation. It is then convenient to expand $\mathfrak R$ in terms of $e^{\lambda_i-\lambda_j}$. The same expansion turns out to be useful also for the metric. \\

We note that the derivatives with respect to the radial variables yield
\eq{\left(\partial_{r^\alpha} p\right)  p^{-1} = k \left(\partial_{r^\alpha} h\right) h^{-1} k^T =k H_\alpha k^T \doteq  X_i \mathfrak R^i_\alpha,}
with $H_\alpha \in \mathfrak h$. Thus, similarly to the $AdS$ cases, the radial components $\mathfrak R^i_\alpha$ do not depend on the radial variables at all and can be neglected. Of the full $8 \times 8$ matrix we therefore need only the components $\mathfrak R^i_j$ with $i=1 \ldots 8$ and $j= 3 \ldots 5$. The expression for the matrix is unfortunately still quite complicated and it would be of little use to explicitly write it down here. Instead we refer to a Mathematica notebook where all of the computations are readily available\cite{arponen.nb}. Fortunately the asymptotic form of the metric simplifies greatly and can be written as
\eq{ds^2 \sim  e^{2 r^1+r^2}(\theta^1)^2 + e^{r^1+2r^2}(\theta^2)^2 + e^{r^1-r^2}(\theta^3)^2}
where the $\sim$ means asymptotical equivalence at large $r^i$ and
\eq{
\left\{ \begin{aligned}
\theta^1 &= \frac{dz-x dy+y dx }{1+x^2+y^2+z^2}\\
\theta^2 &= \frac{dx-y dz+z dy}{1+x^2+y^2+z^2}\\
\theta^3 &= \frac{dy+x dz-z dx}{1+x^2+y^2+z^2}
\end{aligned}
\right. .}
The term $\propto e^{r^1-r^2}$ is actually not a leading term at all in the limits $r^1,r^2 \to \infty$ or $r^1 \to \infty$, but we kept it here to stress the observation that the $\theta^i$'s above can be understood as a coframe on $SO(3,\mathbb R)$, i.e. the dual basis to vector fields $v_i$, which may be solved from the requirement $\theta^i (v_j) \equiv \delta^i_j$ explicitly:
\eq{\left\{ \begin{aligned}
v_1 &= (y+x z)\partial_x + (-x +y z)\partial_y + (1+z^2)\partial_z\\
v_2 &= (1+x^2)\partial_x + (x y+z)\partial_y + (-y + x z)\partial_z\\
v_3 &= (x y-z)\partial_x + (1+y^2)\partial_y + (x+y z)\partial_z.
\end{aligned}\label{frame.vectors}
\right. }
They form the $\mathfrak{so} (3, \mathbb R)$ Lie algebra. However, as we will see below, they cannot all be simultaneously asymptotic symmetries.\\

It can be shown that the boundary of $P(3, \mathbb R)$ is topologically $S^{4}$ (see Appendix 5 of Schr\"{o}der\cite{schroeder}). However, as the asymptotic metric loses all measure of the distances $dr^i$, it is perhaps more reasonable to first consider fixed $r^i$, in which case we are really looking at a subset $S^3 \subset S^4$ (the $S^4$ case will be included in a more complete investigation in the future). We consider first the case $r^1 = r^2 \to \infty$ and define the asymptotic metric as
\eq{d \bar s^2 \doteq  (\bar\theta^1)^2 +(\bar\theta^2)^2 \label{asymptotic.double}
}
with
\eq{
\left\{ \begin{aligned}
\bar\theta^1 &\doteq dz-x dy+y dx \\
\bar\theta^2 &\doteq dx-y dz+z dy \label{asymptotic.forms}
\end{aligned}
\right. .
}
We have also discarded the factors $(1+x^2+y^2+z^2)^{-1}$, since we are interested in what are essentially conformal transformations of the above metric, analogously to the AdS case in eq. (\ref{adsmetric}). It is indeed useful to compare the above asymptotic metric to the standard two dimensional Euclidean metric $dx^2 +dy^2$ (or $d \Omega$). As opposed to the one forms $dx, dy$ the $\bar\theta^i$ (or $\theta^i$) are neither exact nor closed. The metric is therefore not some trivial redefinition of a two dimensional metric.\\

In the single limit case $r^1 \to \infty$ we would just obtain the asymptotic metric

\eq{d \bar s^2 \doteq  (\bar\theta^1)^2.}
Note that only at most two of the one forms $\theta^i$ can be simultaneously present in the asymptotic metric. This is a simple consequence of the tracelessness of the defining representation.

\section{Asymptotic symmetries}
Asymptotic symmetries are here defined as vector fields
\eq{\widehat X \doteq  \xi^1 \partial_x + \xi^2 \partial_y + \xi^3  \partial_z,}
where the components $\xi^i$ depend on $x,y,z$. They transform the metric of eq. (\ref{asymptotic.double}) up to a positive function, i.e.
\eq{ \mathcal L_{\widehat X}  d \bar s^2  = 2 \bar\theta^1 \mathcal L_{\widehat X}  \bar\theta^1  + 2 \bar\theta^2 \mathcal L_{\widehat X}  \bar\theta^2 \equiv \lambda_1 d \bar s^2}
for some unknown function $\lambda_1$. Here $\mathcal L_{\widehat X}$ is the the Lie derivative. We find the conditions
\eq{
\left\{ \begin{aligned}
\mathcal L_{\widehat X}  \bar\theta^1 &= \lambda_1 \bar\theta^1 + \lambda_2  \bar\theta^2 \\
\mathcal L_{\widehat X}  \bar\theta^2 &=  - \lambda_2 \bar\theta^1 + \lambda_1  \bar\theta^2
\end{aligned}
\right. .}
Using the explicit formula for a Lie derivative of a one form,
\eq{\mathcal L_{\widehat X} \theta = \left( \xi^j \partial_j \theta_i + \partial_i \xi^j \theta_j \right)d\omega^i}
and the definitions (\ref{asymptotic.forms}) we obtain six determining equations for the unknowns $ \lambda_i, \xi^i $. There are no derivatives of the $\lambda^i$ so they are easily eliminated and we are left with four equations for the three unknown $\xi^i$. The equations are still rather involved and finding solutions does not seem like a simple task, although it is easy to find some special solutions, which may then be commuted with each other to find new ones. For example, the $v_3$ in eq. (\ref{frame.vectors}) is an asymptotic symmetry as is a simultaneous scaling of $x$ and $z$.  The main body of the computations has been relegated to appendix \ref{appendix.solving} and we collect the final results here.\\

First of all, there is a six dimensional subalgebra of the form $\mathfrak{sl}(2,\mathbb R) \oplus \mathfrak{sl}(2,\mathbb R)$ generated by
\eq{\left\{ \begin{aligned}
l_{-1}  &= \left( 1+ i y\right)\left( \partial_x - i \partial_z \right) \\
l_0  &=   \left( 2x+ i (x y+z) \right)\partial_x + i (1+y^2)\partial_y  +\left( 2z+i (y z-x) \right)\partial_z  \\
l_1  &=   x \left( x+i z \right)\partial_x +  \left( x y-z+ i(x+y z) \right)\partial_y +z \left(x+ i z\right)\partial_z
\end{aligned}
\right. \label{sl2}}
and another copy $\{ \bar l_i\}$ that can be obtained from the algebra above by complex conjugation. It is interesting to note that the maximal subalgebra is of this form given that the algebra of isometries is $sl(3,\mathbb R)$. Note that the above subalgebra is not the algebra of asymptotic \emph{isometries}, since they do not all leave the asymptotic metric completely invariant. Indeed, it is known that the isometries on the boundary act as $so(3,\mathbb R)$ \cite{schroeder}. There are also two independent infinite dimensional Lie algebras generated by
\eq{&&L_F \doteq \frac{(i+y)(x-i z)}{x+y z}F\left( \frac{i+y}{x-i z} \right) \nonumber \\ &&\times  \left[ (x+i z)\partial_x+(-i+y)\partial_y \right] \label{L.generators}}
and a "complex conjugate" with $F$ replaced by $\bar F\left( \frac{-i+y}{x+i z} \right)$. Here the function $F(z)$ is an arbitrary (e.g. meromorphic) function and could be expanded in Fourier modes $\propto e^{i n z}$ to produce the Witt algebra. Lastly, there is yet another infinite dimensional Lie algebra generated by
\eq{&&V_G \doteq G(x,y,z)\nonumber \\ &&\times \left[ (x y-z) \partial_x+(1+y^2)\partial_y+(x+y z)\partial_z) \right]\label{V.generators}}
where $G$ is also arbitrary. We have summarized the Lie algebra satisfied by these generators in the form of a commutator table in Table (\ref{commtable}) below.\\
\begin{table}[ht]
 \caption{Commutator table for the infinite dimensional Lie algebra of asymptotic symmetries of $SL(3)/SO(3)$. Here $\mathfrak l= \{l_{-1},l_0,l_1\}$ and $\mathfrak L \doteq \left\{ L_F \right\}$ denotes all vectors of the form in eq. (\ref{L.generators}) and similarly $\mathfrak V \doteq \left\{ V_G \right\}$. Then the meaning of e.g. the commutator $\left[ l, L_F \right] = V_G$ is that the commutator of a vector in $\mathfrak l$ with a vector of type (\ref{L.generators}) is a vector of type $V_G$ for some function $G$, which we write $\left[ \mathfrak l, \mathfrak L \right] \subseteq \mathfrak V$.\label{commtable}}
\begin{center}
  \begin{math}
  \begin{array}{c||c|c|c|c|c}
          &   \mathfrak l     &  \mathfrak L     & \bar{\mathfrak l}    & \bar{\mathfrak L}   &\mathfrak V  \\
    \hline\hline
  \mathfrak l       &   \mathfrak l     &  \mathfrak V     &     0     &\bar{\mathfrak L}+\mathfrak V  & \mathfrak V  \\
  \mathfrak L       &  \mathfrak V      &   \mathfrak L    & \mathfrak L+\mathfrak V      &   0      &\mathfrak V  \\
    \bar{\mathfrak l}    &   0      & \mathfrak L+\mathfrak V    & \bar{\mathfrak l}    &  \mathfrak V      & 0  \\
    \bar{\mathfrak L}    & \bar{\mathfrak L}+\mathfrak V &    0    &    \mathfrak V     & \bar{\mathfrak L}   &\mathfrak V  \\
   \mathfrak V         &  \mathfrak V      &   \mathfrak V    &     0     &   \mathfrak V     &\mathfrak V
 \end{array}
  \end{math}

\end{center}
\end{table}

It would probably serve no purpose to expand the vectors in a basis at this stage. To the best of my knowledge, the above infinite dimensional Lie algebra is not of any well known type. Also, I cannot guarantee that the above generators are all the solutions of the determining equations, but the above Lie algebra certainly is, at the very least, a particularly large subalgebra.\\

The single limit $r^1 \to \infty$ is a much simpler case, since we only need the asymptotic symmetry of $\bar \theta^1$. It is easy to verify that the general solution is
\eq{V_\Lambda' &&\doteq (x \Lambda _z+\Lambda _y) \partial_x +(y \Lambda _z-\Lambda _x) \partial_y \nonumber \\ &&-(y \Lambda _y+x \Lambda _x-2 \Lambda)\partial_z ,}
where $\Lambda=\Lambda(x,y,z)$.
\section{Concluding remarks and future perspectives}
We have found a five dimensional metric space which admits an infinite dimensional Lie algebra of asymptotic symmetries. What this means is that the metric on the boundary tends to a conformal structure, which is invariant under an infinite dimensional Lie algebra. Such infinite symmetries were previously known only for the two and three dimensional Anti-de Sitter spaces.\\

It is reasonable to ask whether it would be possible to generalize the present results to full fledged gauge or gravity theories with asymptotically $P(3,\mathbb R)$ boundary conditions, or indeed to a corresponding $P(N,\mathbb R)$ case, as has been done previously in three dimensional Chern-Simons theories\cite{henneaux2,campoleoni.theisen,banados} with $AdS$ boundary conditions (note however that e.g. general relativity with a negative cosmological constant is not a viable candidate, since it will be asymptotically $AdS$). In these theories the action of the gauge field is an integral of the Chern-Simons 3-form,
\eq{S \propto \int \tr \left( A \wedge dA + \frac{2}{3}A\wedge A\wedge A\right).}
For $P(3)$ -type boundary conditions one might use instead the Chern-Simons 5-form and the action
\eq{S \propto  \int \tr \left( dA dA A + \frac{3}{2} dA A^3 +\frac{3}{5} A^5 \right),}
where we have dropped the wedges for conciseness. Note that the Maurer-Cartan form $dp p^{-1}$ considered in this work is a solution of the resulting Euler-Lagrange equations of the above action. Such five dimensional Chern-Simons theories are of course not new. However, no results seem to exist with $P(3)$ -type boundary conditions, which may explain the absence of infinite symmetry on the boundary. It is of course far from guaranteed that such models would possess infinite dimensional asymptotic symmetries.\\

Another viable direction is to consider theories covariantly coupled to a fixed background metric/gauge field of type $P(3)$ analogously to the AdS/CFT correspondence. Infinite symmetry near the boundary then follows by construction.

%
%
%
%
%
%
%
%
%
%
%
%
%
%
%
%
%
\appendix
\section{Solving the determining equations\label{appendix.solving}}
The determining equations are
\eq{
\begin{array}{c}
\left\{ \begin{aligned}
 \xi ^2-\lambda _2+y \left(-\lambda _1+\xi _x^1\right)-x \xi _x^2+\xi _x^3 &= 0 \\
 -\xi ^1-z \lambda _2+y \xi _y^1+x \left(\lambda _1-\xi _y^2\right)+\xi _y^3&= 0 \\
 -\lambda _1+y \left(\lambda _2+\xi _z^1\right)-x \xi _z^2+\xi _z^3&= 0 \\
 -\lambda _1+\xi _x^1+z \xi _x^2+y \left(\lambda _2-\xi _x^3\right)&= 0 \\
 \xi ^3-x \lambda _2+\xi _y^1+z \left(-\lambda _1+\xi _y^2\right)-y \xi _y^3 &= 0 \\
 -\xi ^2+\lambda _2+\xi _z^1+z \xi _z^2+y \left(\lambda _1-\xi _z^3\right)&= 0
\end{aligned}
\right. ,
\end{array}
}
where for the sake of conciseness we use subscripts to denote partial derivatives, e.g. $f_x = \partial_x f$ etc. The $\lambda^i$ can be eliminated and we are left with the four equations
\begin{widetext}
\eq{
&\left(y (x y-z)\partial_x+y\left(1+y^2\right)\partial_y+(x+y z)y \partial_z-1+y^2\right)\xi ^1 \nonumber \\ &+\left(x (-x y+z)\partial _x
  -x \left(1+y^2\right)\partial _y-x(x+y z)\partial _z+x y-z\right)\xi ^2 \nonumber \\ &+\left((x y-z)\partial _x+\left(1+y^2\right)\partial _y+\partial _z\right)\xi ^3 =0, }
\eq{&\left(1+y^2\right)\left(x\partial_x +y\partial _y-1\right)\xi ^1 +\left((z(x+y z)-x(x y-z))\partial _x-x \left(1+y^2\right)\partial _y+x y-z\right)\xi ^2\nonumber\\ &+\left(1+y^2\right)\left(-z\partial _x+\partial _y\right)\xi ^3=0,}
\eq{&\left(-y \left(x^2+z^2\right)\partial _x+x\left(1-y^2\right)\partial _y+2 y z\partial _z+x y-z\right)\xi ^1+\left(x^2+z^2\right)\left(x\partial _x+y\partial _y-1\right)\xi ^2\nonumber\\ &+\left(-\left(x^2+z^2\right)\partial _x+(z-x y-y (x+y z))\partial _y+x+y z\right)\xi ^3=0,}
\eq{\left(y\partial _x+\partial _z\right) \xi ^1+\left(-x\partial _x+z\partial _z\right)\xi ^2+\left(\partial _x-y\partial _z\right)\xi ^3 =0 .}
\end{widetext}
A common strategy for solving determining equations of this type is to transform the system into a general canonical form by e.g. considering an equivalent system of polynomial equations, which is rewritten in a Groebner/standard basis (see e.g. chapter 10 of Baumann\cite{baumann}). Such computations may be delegated to a sophisticated computer algebra software, such as the "Rif" package which is built in the latest versions of Maple, or for example the commercial Mathematica package MathLie\cite{baumann}. However, neither of the above mentioned systems are able to solve the above determining equations outright. It is still possible to obtain (probably all of the) solutions by combining the computer methods with some human intuition. Verifying that a vector field satisfies the determining equations is luckily easy for example by using the author's Mathematica notebook\cite{arponen.nb}. As mentioned in section 2, the vector $v_3 \doteq (x y-z)\partial_x + (1+y^2)\partial_y + (x+y z)\partial_z$ of eqs. (\ref{frame.vectors}) is a solution, as is the scaling $x \partial_x + z \partial_z$ and e.g. $\partial_x + y \partial_z$. Similar guesswork and commutations of found generators lead to an algebra generated by
\eq{
\left\{
\begin{aligned}
u_1 &= x \partial_x + z \partial_z \\
u_2 &= \partial_x + y \partial_z \\
u_3 &= y \partial_x -\partial_z \\
u_4 &= (1+x^2)\partial_x +(x y-z)\partial_z +(x z+y)\partial_z \\
u_5 &= (x y+z) \partial_x + (1+y^2)\partial_y+(y z-x)\partial_z \\
u_6 &= (x z -y) \partial_x +(x+y z)\partial_y + (1+z^2)\partial_z
\end{aligned}
\right.}
Note that the generators ${u_4,u_5,u_6}$ are not the same ones as in eq. (\ref{frame.vectors}). This Lie algebra is in fact $\mathfrak{sl}(2,\mathbb R) \oplus \mathfrak{sl}(2,\mathbb R)$, and after a change of basis, the generators are expressed as in eq. (\ref{sl2}).\\

The determining equations will turn out to be easier to analyze if we attempt a gradation of the Lie algebra with respect to an element in the Cartan subalgebra, for example the scaling generator $u_1$ above. Thus we demand $[ u_1 , \widehat V ] = n \widehat V$ for some integer $n$ (noninteger values would lead to nonanalyticity), resulting in three additional equations
\eq{
\left\{
\begin{aligned}
  x \xi^1_x + z \xi^1_z  &= (n+1)\xi^1 \\
  x \xi^2_x + z \xi^2_z  &= n\xi^2   \\
  x \xi^3_x + z \xi^3_z  &= (n+1)\xi^3.
\end{aligned}
\right.}
For example if $n=0$, we find the solutions
\eq{\left\{
\begin{aligned}
  \xi^1 &= \frac{x(x y-z)}{x+y z} f(y,z/x) + c_1 \frac{y(x^2+z^2)}{x+y z}+c_2 \frac{x^2+z^2}{x+y z} \\
  \xi^2 &= \frac{x(1+y^2)}{x+y z} f(y,z/x) + c_1 \frac{x(1+y^2)}{x+y z}-c_2 \frac{z(1+y^2)}{x+y z} \\
  \xi^3 &= x f(y,z/x)
\end{aligned},
\right.}
where the $c_i$ are free constants and $f$ is a free function. We can also find special solutions similar to the $c_i$ above for any $n$, which turn out to be exactly the $L_F, \bar L_F$ in eq. (\ref{L.generators}). Commuting the $l, \bar l$ with $L_F, \bar L_F$ produces vectors of type $V_G$ of eq. (\ref{V.generators}) and we have therefore found a closed infinite dimensional Lie algebra as summarized in table (\ref{commtable}).
%
%
%
%
%
\bibliographystyle{hunsrt}
\bibliography{bibsit2}
\end{document}